# Aesthetical Attributes for Segmenting Arabic Word

Mohamed Hssini[1], Azzeddine Lazrek[2]

Department of Computer Science, Faculty of Sciences, Cadi Ayyad University
Marrakech, Morocco
[1]m.hssini@ucam.ac.ma; [2]lazrek@ucam.ac.ma

*Abstract*- The connected allograph representing calligraphic Arabic word does not appear individually in any calligraphic resource but in association with other letters all adapted to each other. The graphic segmentation of the word by respecting aesthetical attributes indicating the grapheme of every letter is far from being an obvious task. The question consists in discovering every letter constituting the word, points of cutting which separate its grapheme from other constituents of word's shape. The obtained segment must be a complete drawing of the represented letter. This segmentation according to contextual graphic and qualitative criteria connecting the attached allograph will have to satisfy typographic constraints varying in conformity with the possibilities offered by the wanted technology. In this paper, we develop an approach for segmenting Arabic word from which the purpose is to extract graphemes respecting the design of Arabic letters such as it is in the calligraphic literature. The procedure bases itself on the principle that the Arabic connected letters have a common part included in the cursive area, which must not be lost during the process of cutting.

*Keywords- Typographic Modeling; Arabic Calligraphy; Graphic Quality; Cursive Area; Procrustes Method; Digital Font*

## I. INTRODUCTION

The typographic modeling of the writing is based on the Gutenberg approach which consists in subdividing it in typographical unit signs. At the beginning, this process was perfected for visual use, aiming to produce Latin script with the most feasible fidelity. The graphemes representing Latin letters lend themselves easily to dissociate; they are distinguishable without difficulty. However, this feature is not shared by a cursive writing with calligraphic patterns, such as the case for Arabic script that involves more than just lining up letters. In Arabic, the attached letters take on board with each other. They are exceptionally adaptable, which makes it less easy to illustrate each allograph independently. In calligraphic resources, we never find the individual context allograph for any Arabic letter. The concept of discrete variant of letter's grapheme as in digital typography did not exist in Arabic calligraphy [1].

It follows an incompatibility between the Gutenberg model and the graphical properties of Arabic script. Typographers have been debating for a long time to treat this problem. The known attempts led to resolving this difficulty based on some typographic simplifications [1]. This incompatibility would be omitted by according Arabic script a structure one and the same to that of Latin script. The choice of unit segments of typographical traditional model of writing cannot be separated from technological advances developed at that time at the beginning of the printing era. The arrival of the computer has given a new opportunity to appropriate a typographical model closer to graphical properties of Arabic. The question of authenticity of this script arises again including the segmentation of calligraphic word which is the first step in this way.

Segmenting calligraphic word is not only a subject which is a part of atomicity of digital text, but it is a framework that helps for extracting information concerning the internal relationships in calligraphic word. Then, the process of segmentation must illustrate the complete grapheme of each constituent of the word. The importance of this subject is that the word is an object of study of various fields. Seen these constraints, the graphic segmentation of Arabic word is far from being an obvious task. The problem consists in finding, for every letter constituting the word, a region of candidate cutting points which separates its whole grapheme from the other constituents of the shape's word; according to contextual graphic and qualitative criteria connecting the attached allographs. This region must be linked to two segments representing each letter.

The obtained segments will have to satisfy typographic constraints varying according to the possibilities offered by the wanted technology, if they have been oriented to be glyphs in digital font.

The cut graphemes serve to study the variability of the shape of a letter. To serve as glyphs for a digital font, normalization and search for an average shape have to take place.

In this paper, we present a segmentation procedure of the word by basing itself on calligraphic rules, and the various issues which ensue during the implementation of the graphic cutting of the word in the preparation of digital fonts with calligraphic glyphs.

After a reminder of the properties of the Arabic writing and the describing of the cursive relations in the Naskh style in Section 2, we try to present an overview of modalities for segmenting Arabic word in Section 3. Section 4 is devoted to presenting the design of cursive area. Section 5 is consecrated to study the contextual varieties of the cursive area, the main effects of the justification on this surface. Section 6 presents the main criteria to measure the graphic quality of the area to be cut. The method of segmentation makes object of study in Section 7. The connected problems to an implementation make object of Section 8. Eventually, we end with conclusion and perspectives.

II.  GRAPHICAL SEGMENTATION OF ARABIC WORD

*A. Proprieties of Arabic Writing*

  *1) Arabic Alphabet:*

The Arabic writing is characterized by its cursivity. There is a complex set of rules that govern the connections of letters and the design of their diacritics [2]. The Arabic alphabet contains twenty-eight fundamental letters. Every letter - with the exception of a number of six letters ا ر ز د ذ و - becomes attached to the others. The Arabic letters possess till four base forms in a precise context dependent on the letters and letter's place in the word: in an isolated, initial, median or final position within a word. Besides, according to the nearby letters or other contexts, every base form can have several allographs [3, 4].

In the Arabic writing, several letters have identical plans and are distinguished only by the number of points, registered or signed according to the number of ambiguities to remove. Points also work to spread the number of possible graphemes. The point has a lozenge shape. It is obtained by pressing the pen diagonally across the paper. The four equal sides of the point being the size of the width of the pen.

The point has an additional role. It serves as a tool and unit of measure (see Fig. 1). It involves determining the length, width, depth and degree of inclination of the letter.

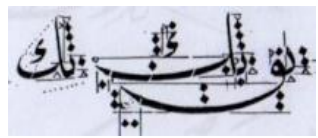

Fig. 1  Measuring by point

  *2) Diacritical Marks:*

The Arabic diacritical marks are necessary to differentiate certain letters or certain words, and for aesthetic purposes [2, 5].

  *a) Role*

Arabic diacritical marks aim to:

- Readability;
- Changing the phonetic value of a letter;
- Avoiding any ambiguity between two homographs;
- Determining the grammatical role of the word;
- Filling the blank in the word;
- Playing an aesthetic role.

A word such as "شعر" when is diacritized can be: "شَعَرَ" he sense, "شِعْر" poetry, "شَعْر" hair.

  *b) Classification*

There are three types of Arabic diacritical marks, in conformity with their typography and their grammatical role [2, 4, 5]:

- Language's diacritics:

Language's diacritics determine the meaning of word (see Fig. 2). They come into sight as:

  - Diacritics above: positioned above a basic letter, as Fatha, Damma, Sukun, or Shadda.
  - Diacritics below: positioned below a letter, as Kasra or Kasratan.

- Aesthetic diacritics: often fill space created when extending some letters, to improve the aesthetics.
- Explanatory diacritics: placed to differentiate the letters with or without dots.

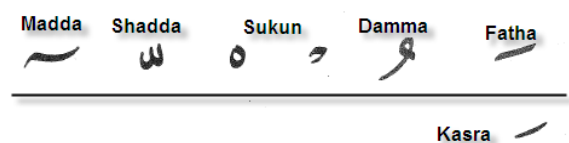

Fig. 2  Some Arabic diacritics

  *c) Sizes*

Every Arabic diacritic is characterized by its length, its width, its shape and its angle of curvature. These dimensions are measured, as those of the letters, by the point.

The two diacritics Fatha and Fathatan can be elongated in some contexts. The size of these diacritics depends on the size and height of its base letter, and on those neighboring base letters and positions and sizes of their diacritics [2, 5].

*d) Baseline*

The baseline is an imaginary line, which regulates the shape of basic letters and the junction of the words.

The Arabic letters are characterized by their position in front of the baseline. According to the context, there are:

- letters on the baseline, as: ا-ب-ط-ف-ك-ل-م-ه-لا
- letters which have a part under the baseline, as: ن-ص-ل-ى-س ر-و-ق-ح-م-ع

*e) Diversity of Styles*

There is a great variety of Arabic calligraphy styles that vary by region, time, materials and calligrapher. To mention some of them: Kufic, Naskh, Nastaliq, etc. Each style has its features, and follows its sub schools. Each school gives its own mark to the style. Arabic styles differentiate principally by [6]:

- Geometric shape of the letters;
- Number and shape of letter variants;
- Presence, shape and number of dots;
- Presence, shape, role and number of diacritics;
- Use, shape and size of Kashida.

Each writing style has its own strict rules and context (edition, illustration, architectural decoration, etc.). This study concerns only Naskh style.

*3) Ligatures:*

Some attached letters change the positions with forms and they are stacked vertically. The layout of this set is intertwined and does not rigorously follow the baseline. For that, the typographic approach is to represent this set of letters by ligature in which these letters are combined forming one character. There are three kinds of ligatures: contextual, linguistic and aesthetic. The contextual ligatures are needed for cursive writing.

*4) Cursive Connections:*

*a) Letters Families*

In isolated form, 18 Arabic letters differ only by the presence or absence of points, as well as their position and their number if necessary. Every calligraphic rule applied to a letter applies also to the other letters of the same family. The groups or the families of letters are distributed as following [6]:

{ب ت ث}=[ب]

{ح ج خ}=[ح]

{د ذ}=[د]

{ر ز}=[ر]

{س ش}=[س]

{ص ض}=[ص]

{ط ظ}=[ط]

{ع غ}=[ع]

Then the single letters: ا ف ق و ه ي ن م ك ل.

*b) Allograph*

In cause of the cursive nature of the Arabic writing, the Arabic letters possess variants said allographs. It is the diverse graphic forms which a letter can have while keeping its place: isolated, initial, median or final. For every graphic environment, and according to the diverse contexts, each Arabic letter has a graphic shape while guarding its square in the word.

*5) Nature of Connections:*

It follows that the nature of letters connections differs from one context to another:

- The family [ح] (see Fig. 3) and the letter م cause multi-lines in the word. Certain letters are obliged to adapt their forms in front of contexts producing the multi-line;

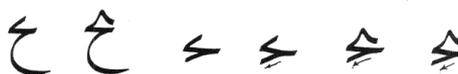

Fig. 3  Connectivity with [ح]

- Connections with the letter ص: The letter ص is linked to other letters with cursive area in laying shape (see Fig. 4);

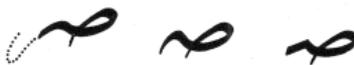

Fig. 4  Connectivity with ص

- Letters with teeth (see Fig. 5);

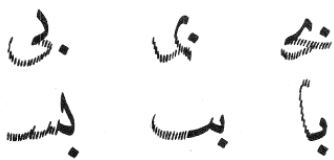

Fig. 5  Letters with teeth [ب] in cursive state

- Reducing the size of some letters: Certain letters are obliged to adapt their forms in front of contexts producing the multi-line (see Fig. 6);

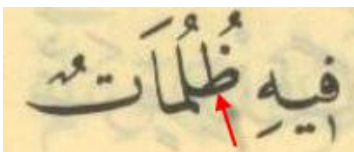

Fig. 6  Adapting the form of letter ظ

- Letter ك is linked with adjacent and non-adjacent letters (see Fig. 7);

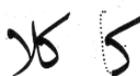

Fig. 7  Double connectivity with letter ك

- Connections with the letter ل (see Fig. 8).

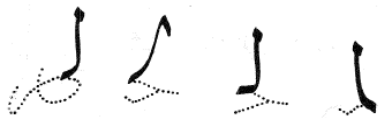

Fig. 8  Some allograph of letter ل

*6) Word and Justification:*

In the Arabic writing, a word can be dilated by the Kashida to cover much space, and can be pressed by the use of the ligatures. These mechanisms could influence the sizing and the positioning of the Arabic diacritical marks [2, 5]. Therefore, some of the used tools are [8, 9]:

- Kashida: it occurs for some letters, according to some situations, following some conditions, stretching in some sizes;
- Ligatures: they occur where letters are superposed; they are seen as tools of contracting Arabic words;
- Moving the final letter for contracting latest word in line.
- Size reduction of last letter;
- Allograph use, where glyph is substituted by another;
- Accumulation of words;
- Writing in the margin.

The optimum fit algorithm - applying in $T_EX$ system - has been applied to Arabic justification by taking into account the variants of glyphs given by the jalt table in Open Type format [9].

*B. Graphic Segmentation of Word*

*1) The Arabic Word:*

The word is a textual unit defined by the typesetters. Graphically, the Latin word is bordered by space, apostrophe or hyphenation.

In Unicode, a word's shape is a sequence of characters delimited by two boundaries words.

In Arabic script, the word consists of a combination of:

- Subwords: a subword is a connected part of the word which consists of letters attached to each other. It therefore has a letter in the initial form, then medians letters and ends with a letter in final form;
- Isolated letters.

These subwords contained in the word end with letters which have only two contextual forms, or by letters which have four contextual forms and which are at the final contextual form in the word end.

*2) Arabic Word Boundaries:*

More than the identification of word, there are many uses of word boundaries in a number of areas. We count for examples: database queries, mouse selection, searching, and cut and paste.

For the Arabic language, Unicode has restricted Arabic word boundaries to whitespace and punctuation. However, these two factors do not always separate two neighboring words in Arabic calligraphy.

The isolated letters and the diacritical marks are detached from the rest of the graphic word. However, the isolated letters can be interwoven. This phenomenon takes place at the end of the word and before the followed word in the same line, without spaces which separate them. The interweaving takes place in very precise and specific contexts.

The couples of letters forming an interweaving, depending on the Calligrapher Mohamed Chawki, are presented as follows (Fig. 9): for every family of letters with final or isolated state, we decipher the letters which form with them a context of interweaving:

- The family [ب] is interwoven with the letters: ا] ب] ن س ل;
- The family [د] is interwoven with the letter: ا;
- The family [ر] is interwoven with the letters: و ا;
- The letter ن is interwoven with the letters: [ح] م ا ل;
- The letter ل is interwoven with the letters: [س] [ح] ا;
- The letter ي is interwoven with the letters: [س] ا.

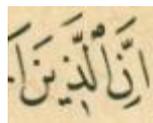

Fig. 9  Interwoven letters

*3) Graphic Segmentation of Word:*

The graphical segmentation is a dissociation of the graph illustrating the word into graphemes describing each of its components. The desirable separation must bring information and allow us the good exploration of the internal calligraphic relationships contained in the representation of the word. However, in calligraphic exercises writing, the contextual connected allograph does not appear individually but in association with other letters all adapted to each other. Both attached letters have, following the context, a common part which is in a zone of cursivity; where the exit of the first one and the entry of the second are matched together. The shape depends on the allographic context formed by both attached letters, sudden effects of the justification by the word, limitrographs and graphic quality of the graph to be cut. In this way, segmentation is a process of decontextualization causing a loss of contextual and graphic information. The choice of the cutting points must be based on criteria strengthened by a microallographic investigation in the contour of the zone to be separated.

III. MODALITIES FOR SEGMENTING ARABIC WORD

*A. Graphic Segmentation in Digital Typography*

*1) Arabic Graphic Segmentation and Print:*

The typographic approach which results from the Gutenberg model is to represent writing by segmenting it in individual typographic signs. The constituent unit was the metal block.

Generally, the technique of printing was not able to offer the possibility to control the graphic aspects of the Arabic text with extra level of authenticity: adapting grapheme of each Arabic letter to constraints imposed by the support, controlling the body and the position of allograph into the cursive attachment and justification, managing the contexts for freely associate letters and diacritics, and so on [10]. Continuous technological developments that typography had undergone during different centuries have not changed at the bottom of the Gutenberg model. The invention of the keyboard has considerably accelerated the process of composition and simplified the interaction between human and machine. This step paved the way to automate the typographic process, but has nothing added to solve the problems of the Arabic script [10].

Bearing in mind the mentioned constraints, the Arabic typographers come to the conclusion that the problem to solve is more general, and that the solution had to be a simplification of the Arabic script. Arabic typography has introduced new graphic variables to represent the basic letters and diacritics. The reform for adapting Arabic writing to typographic process mainly aims to solve one principal technical problem [1]: minimizing the number of types needed for typesetting Arabic writing.

The first typesetters who were interested in printing Arabic calligraphy were Westerners and have not enough knowledge about rules of Arabic calligraphy. They tried to imitate the handwriting including the most complex connections. The ligatures were a way to approach the complex cursive relationships. However, all the styles of writing are not easily replicable in printing. In a spirit of rationalization, we abandoned the ligatures and we adopted the principle of separation of the characters to limit the number and thus the printing costs [11]. Letters are connected along a linear area. For this reason, the Arabic writing was simplified in mono-line writing.

*2) Atomicity of Digital Text:*

The appearance of new information technologies has revealed new uses for Arabic writing. The state was comparable to the situation at the beginning of Arabic printing. We had to remodel an existing heritage to adapt it to an environment that worked for a new manner [10].

Studying the atomicity of textual data, among which the visual aspect is a part, is an essential step in trying to develop a digital production model for the authentic Arabic script. Improving the quality of text visualization finds in Arabic calligraphy supporting rules.

Document composition system executes a series of textual processing, from a representation of exchange to a final representation, through a transitional internal representation. The process of the exchange composition, the internal handling and storing of a text most often assume that it is represented in the form of strings, while the display of text takes place through a visual model whose textual unit is not the character but the image of it: the glyph.

Typographical concepts have undergone many changes. Initially the word character referred to engraved mark. Afterwards it gained new significances including: individual sign, individual block of metal, or the entire font, that is to say all signs which share a common graphic. This is the meaning of *individual sign* of the word character that was included in the digital typography. It was not illogical to call the atomic element character text in computers and all characters available on a given platform, with their numerical values, the character encoding or character set. In Unicode, character description is simple and logical or linguistic dominance of an equivalence class of glyphs. A glyph is the image of a typographical sign as it is contained in a digital font. Some characters have several glyphs, and conversely, some glyphs represent more than one character [10].

In digital typesetting, for against, the concept of font has undergone many changes. It is a set of correspondence between character codes and glyphs. It should be noted that the control characters, also called non-printing characters, do not have dedicated glyphs.

The digital textual models generating the Arabic writing are ensued from the Arabic typographic textual models before the era of computers. Several calligraphic properties, which particularize Arabic, were simplified to improve a printed model adapting Arabic to the material constraints of the printing technology.

*3) Digital Fonts:*

Digital fonts are essential resources for any system of composition. It is a discreet set of characters of the same family tidied up by size [12].

There are two big types of fonts:

- Bitmap fonts: bitmap font is a set of glyphs which are described by binary matrices. Advantages of bitmap fonts include:
  - Extremely fast and simple to render;
  - Unscaled bitmap fonts always give exactly the same output;
  - Easier to create than other kinds.

The primary disadvantage of bitmap fonts is that the visual quality tends to be poor when scaled or otherwise transformed, compared to outline and stroke fonts, and providing many optimized and purpose-made sizes of the same font dramatically increases memory usage.

- Vector fonts: their principle is based on a mathematical description of the outline of every glyph. The graphic representation of fonts resorts to the curves of Bézier.

The curves of Bézier are polynomial curves described for the first time at 1972 by Pierre Bézier, who used them to conceive by computer, automobile cars. The most important curves of Bézier are the cubic, which are used in computing in the formats of fonts PostScript, Metafont to draw curves "ribbands" joining points or polygons of Bézier. Fonts TrueType use curves of simpler quadratic Bézier.

One of the tasks of fonts is to establish the connection between the textual data of input and those of the visual presentation of the text by the conversion of characters codes to glyphs codes [12, 13].

*4) Arabic Fonts:*

The digital composition of the Arabic text bases itself on the same simplifications followed in the printing model. These simplifications give Arabic writing a structure similar to that of the Latin script.

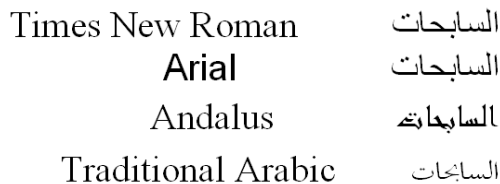

Fig. 10　　Arabic fonts rendering

Some remarks (see Fig. 10):

- Few fonts belong to calligraphic glyphs.
- The segmentation of the word graph is done without studying the variations microallographic.
- The segments representing the letters are changed.
- The connections are linear.
- Few of allographs.
- …

The Arabic fonts represent correspondences between characters and variants of their glyphs. However, in other non-standard environments, there is a different textual granularity. In the Kelk and Imad software, which are systems of layout that offer improvements, fonts work on the combination of some segment sub-letters.

The atomicity of Arabic writing is not different from other alphabetic scripts, but care has to be taken to leave its graphical structure intact. There are some calligraphic rules that should be taken into account in the computer processing of word segmentation. In typesetting, these features have been simplified because of the hardware constraints cited above. Pattern recognition has to be beaten on a deep investigation about these rules.

*B. Graphic Segmentation in Pattern Recognition*

In the theory of pattern recognition, where the segmentation is an indispensable level, two principal strategies are adopted to segment the word in characters [14, 15]: direct segmentation, which consists of decomposing the word into characters; or indirect segmentation, where the word is cut in primitive.

*1) Principles Methods:*

The principal methods used for segmenting Arabic word can be resumed in four techniques, as following [16]:

- Segmentation by Skeleton Analysis: all algorithms following this method segment Arabic word in skeletons.
- Segmentation by Contour Analysis: algorithms adopting this method extract word contours by contour following procedure. They determine the best candidate points of division between graphemes based on the local extreme of the contour, which are associated according to a criterion of proximity.
- Segmentation by Stroke Thickness Analysis: this method use information about the thickness of the tracing for segmenting the word. This estimation is computed by two methods: vertical projections and contour following.
- Segmentation Algorithm based Singularities and Regularities: this method is an adaptation of Simon's algorithm for Arabic handwriting segmentation. First, a pre-processing stage is executed: binarization, text slant correction and then identification and extraction of pseudo-words and secondary parts.

IV. DESIGN OF CURSIVE AREA

It is necessary to designate the attributes on which the attachment of the Arabic letters is based. The design of Arabic letters has its features which distinguish it from other writings. In this section, the aim is to explore these aesthetic features.

*A. Anatomy of Letter*

Studying the structural elements of a graphic representation of a letter allows seeing the path of the letter portions attributing an identity to the hand of the calligrapher. In Arabic calligraphy, there is a set of terms to describe the anatomical structure of a letter. By dividing down letters into parts, we can well again understand how grapheme is created and how Arabic letter is attached cursively.

The difference between the ascending and descending letters which reports the space required for composing the word, define the notion of body. It is a fundamental reference which allows us possessing information on the line spacing and on the space occupied by the writing.

*1) Graphic Letter Portions:*

Letter deconstruction is the first step in learning calligraphy (see Fig. 11). It consists to indicate numbers and arrows of the drawing direction of each part in the graphic representation of the letter. In this sense, dividing the zone cursive between two letters can show interactions between parts of the cursive.

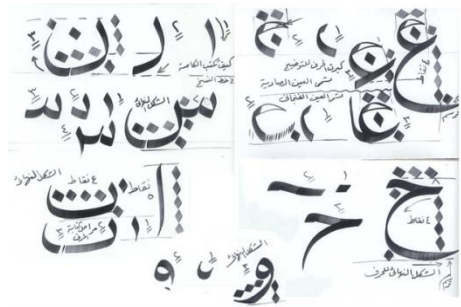

Fig. 11　　Graphic portions of some letters

*2) Regulating Model in Isolated Letter:*

The circle that has the letter Alef as diameter serves as regulating model of the letter's graph in isolated form. The isolated letters are made of graphic parts. A letter draws in report to a direction which determines, according to Ibn Muqlah calligrapher, the beginning and the end of the design of the letter.

There are three types of isolated letter on the report of their beginning:

- The letters which begin with a point: ا [ب] [د] [ر] [س] ل ن ه ;
- The letters which begin with a fragment "Chadia": [ع] ك ي [ح] [ص] [ط];
- The letters which begin with a portion "Jalafa": ف ق و م.

In relation to their end, three types of isolated letters can be distinguished:

- The letters which have an exit with a point: [ب] [د] [ط] ف ك;
- The letters which have an exit with a fragment: [ح] [ر] [س] [ص] [ع] ق ل م ن ه و ي;
- The letters which have an exit with a portion: ا.

*3) Input and Output of Letter in Cursive Forms:*

In cursive state, a letter admits a radical and terminations: one input and/or output.

In the writing of the cursive letters, the hand of the calligrapher is raised only when there is complete exhaustion of the ink.

Let us take an example of two letters to be connected: initial Lam and median Feh, The output of the Lam comes just after the input of Feh: there is a common part between the two connected letters (see Fig. 12).

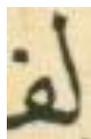

Fig. 12　　Common part between Lam and Feh

The positions of the endings of a median letter are not necessarily in the same position of the basic line. Some conditions must be respected:

- Saturation: having no fractures.
- Completion: every part of the letter must be presented with well specified measures.
- Forms of the parts: every part of the radical or the endings must be respectfully drawn by the shape (curvilinear, linear or circular), measures inclinations of each party.

*B. Design of the Cursive Zone*

The cursive area is the surface which connects both radical parts of both successive letters in the same subword. The length of cursive area varies, in the case of non-extension, between zero and three points. Its shape depends on those two allographs to be connected. The cursive area can be stretched out by Kashida and can reach a length of thirteen points. The calligrapher must respect some conditions for drawing this zone in order to attach the two allographs constituting the context.

*1) Shape:*

The cursive area has five shapes (see Figs. 13 and 14) [17, 18]:

(1) The concave shape;

(2) The linear shape;

(3) The curvilinear shape without curvature;

(4) The curvilinear shape with curvature;

(5) The shape of laying.

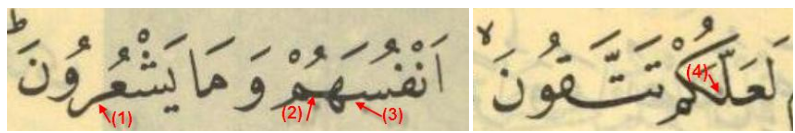

Fig. 13    Shapes of the cursive area except shape of laying

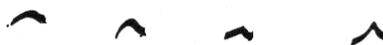

Fig. 14    Cursive area shape of laying

The shape of the cursive area has a link with:

- The allograph of both letters to be attached;
- The slope of the output of the first letter and the input of the second;
- The articulation relative to the baseline;
- The length of the cursive area.

The variation of the length of the cursive area is due to effects of the justification and aesthetic needs. For example, if the number of letters constituting the word is raised, the elongation of the cursive area is necessary for balancing.

As regarding the dimensions, it is a question of determining: the minimal length of the area for a precise context and the thickness of the area.

*2) The Line Thickness:*

The line thickness of writing is a variant of the Arabic writing. It defines, with the body, the main spatial attributes of an Arabic calligraphic style. The thickness of the cursive zone varies following the textual and graphical context (see Fig. 15).

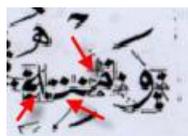

Fig. 15    Variability of line thickness

*3) Cursive Area and Kashida:*

Kashida, as elongation of the cursive area, is present when the area's letter connectivity is stretched out more than three points. The Kashida is characterized by the form that depends on the writing style.

*a) Stretching Places*

Stretching by Kashida occurs in a word according to aesthetics and typographic criteria, and in respecting some roles. For example, it is a defect to superposing two elongations in two consecutive lines.

  *b)  Degree of Extensibility*

The degree of extensibility of stretchable letters depends on some contextual elements:

- nature of the letter to stretch;
- position of the letter in the word;
- position of the word in the line;
- level of writing where Kashida must take place;
- writing style.

  *4) Cursive Area and Ligatures:*

Arabic writing is characterized by multiple baselines, used to position letters in ligatures vertically, known as multilevel or stacking of Arabic writing.

Ligatures introduce the multilevel of writing. The aesthetic ligatures were a fairly limited choice to represent the multilevel of the Arabic script: the number of combinations represented in a multilevel context is large enough that one can guess their representation in a block. This property is caused also by letters: Family Jeem, Meem, and Yeh Barree. Yeh Barree occurs at the end of a word when is preceded by another letter whose glyph ends with the body of a Noon, or Kaas.

  *5) Balance:*

When composing the Arabic word, the calligrapher must fill the void by creating a balance between black and white. This symmetry, depending on our spiritual equilibrium, assists in creating peace within us [5] (see Fig. 16).

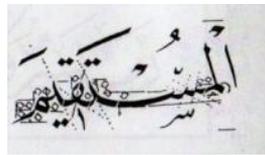

Fig. 16　　Balance between black and white

V. VARIABILITY ANALYSIS OF CURSIVE AREA

Graphic variations of the cursive area are due to changes in mental and physical behaviors of the calligrapher. Modeling the calligraphic production process of the word is a necessity to approach a theoretical framework in which we can study these variations.

*A. Production of the Area Cursive*

The calligraphic representation of Arabic word is the result of two separate processes: mental cognitive process and another physical space-time one.

  *1) Cognitive Process:*

The mental cognitive process manifests itself by a succession of decision states, in which the calligrapher made the choice of covenants allographs in the context. The mental model is a benchmark for the physical realization. It is a simplified approach to the calligrapher intellectual processes [3].

  *2) Physical Process:*

The cursive area is an intermediate passageway between the two radicals of attached letters. The tracing of this path is produced by physical movements of the hand. The calligrapher deformed the Calame beak and exerts a pressure at a certain speed for surrounding the ink flow in order to mastering the physical realization of mental model to draw the cursive area. The Arabic word is a two-dimensional trajectory which grows in a three-dimensional space [3]:

• The position: curved deviations of the letter portions and the changes of the line thickness required to rotate the pen, to lift or push it gradually in order to good drawing the Arabic letter parts.

• The speed: the hand movements must be performed with a certain speed to control the flow of ink between the Calame tip and paper.

• The pressure: it's the force exerted by the hand on the writing tool. The pressure determines the deformation of the Calame to select the beak parts on which to build.

A good path is an indicator of a good movement but is not present that we cannot observe directly because it is a path in a calligraphic historical document.

*B. Contour Extraction*

The contour is the line delimiting the boundary of the calligraphic shape of word represented in digital image. The objective is to explore and comprehend the information offered in the scanned image. The identification of contours is fundamental for analyzing the contents of calligraphic illustration of the Arabic word. For that, the contour extraction is an essential step in this way. The problem is particularly difficult because we are in images case with complex shapes and with noise. Several techniques are used, among which we cite:

- Local methods: examine a small region about every pixel and connect neighboring pixels if they assure some hypothesis.

- Regional methods utilize special techniques to connect pixels which are formerly recognized to be part of the same region or contour.

- Global methods do not rely on any kind of prior knowledge, and they try to find sets of pixels which lie on curves of specific shapes.

Determining the output and the input of calligraphic glyph requires complicated methods which are, unfortunately, to a large degree dependent on the given image material. This problem is especially difficult for images with complex shapes and with noise.

The active contour method is an approach preferably used in the case of very noisy images. Active contours are structures for moving points that we try to change them to marry at best as an interest object. This method consist in moving the points in order to approach them to areas with strong gradient while trying to retain certain characteristics such as curvature of the contour, the distribution of points on it, or other constraints related to available items.

*C. Micro-variability of Cursive Area*

The hand of the calligrapher is characterized by infirm variations that occur between different instances of the same allograph. Patterns of these variations serve to identify features of individual graphic behavior: the movements of the hand, the degree of concentration and mental and physical conditions.

The superposition of the contours extracted from different instances of the same cursive area type allows visualizing:

- The extent of the space in which vary the size and thickness of each type of cursive area.

- The frequency of a graphic portion to be in a given context.

To approximate the surface extent measurement of each cursive area type we must find benchmark points for segmenting cursive area in portions. The forms analysis is based on statistical benchmarks along the contours extracted from multiple instances of the same type. The goal is to determining the frequency of space variation.

TABLE 1 VARIABILITY OF SIZE AND THICKNESS OF CURSIVE AREA

| Type | Variability | |
|---|---|---|
| | Size | Thickness |
| Concave shape | 35% | 33% |
| Linear shape | 28% | 30% |
| Curvilinear with curvature | 43% | 37% |
| Curvilinear without curvature | 34% | 30% |
| Shape of laying | 29% | 31% |

Variability of size and thickness is due to movements that produce the inclinations of the Calame beak over baseline. From the last table (see Table 1) it can be noted that:

- Size and thickness variations are not constant from one cursive area type to another.

- The size variation is larger than the thickness variation.

- The thickness varies depending on the size.

VI. MEASURE OF GRAPHIC QUALITY

*A. Aesthetic Criteria*

In this subsection, we will try to identify the attributes that affect the aesthetics of the area to be subdivided.

*1) Fractures of Portions:*

The junction of the portions can cause fractures. It is observed that fractures of the cursive area can give false information on the candidate points for cutting.

*2) Regularity:*

The regularity of the contour means that it must be continuous. The quality of the graphic can be influenced by a zigzag due to trembling hand, excessive or toner runs out and quality of paper.

*B. Quality of Elongation*

It is a question of controlling degrees and presence of the elongations following the criteria mentioned above.

*1) Distances Between Portions:*

In calligraphic literature, there are distances between the portions of the connected letters expressed in dots (see Fig. 17).

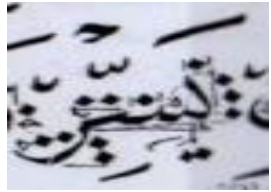

Fig. 17　　Distance between س and ر

*2) Estimation of Graphic Quality:*

This is not only to quantify the extent of regularity but also the global regularity can be approximated by measures of the image entropy [19].

VII. PROCEDURE FOR CUTTING

In this section, we present our segmentation procedure from which the purpose is to extract graphemes respecting the design of the Arabic letters, such as it is in the calligraphic literature. The procedure bases itself on the principle that the Arabic connected letters have a common part included in the cursive area. To separate graphs in two different parts produces a loss of a part of the graph representing at least one of both letters.

*A. Procedure*

- Determining the context.
- Controlling the graphic quality of the area.
- Determining the abnormalities: fractures, false elongations and false decreases of distance between the radicals.
- Neglecting the abnormalities.
- Reducing the cursive area.
- Determining the output of the first letter in conformity with its shape.
- Determining the input of the first letter according to its shape.
- Extracting the common part between both letters.
- Cutting the cursive area in three parts: common part and two others which contain the radicals of letters.
- Duplicating the common part.
- Merging the common part with the parts containing the radicals of letters.

*B. Description*

*1) Determining the Context:*

The contexts which influence the shape, position and dimensions of the length and the width of cursive area, are of three sorts:

- The contexts of neighborhood:
- The elongation is forbidden in some cases: between the letter ب and the family [ح] for example, the cursive area measures zero points.
- The elongation is recommended between two letters with tooth which succeed one another: the cursive area is stretched out by three points.
- The contexts of the limits of the words: the interweaving of the purposes of the word, the letter Yeh Barree and multi-line influence the position of the cursive area to baseline.
- The contexts of the limits of line (effects of the justification).

*2) Controlling the Graphic Quality of the Cursive Area:*

The sources of the anomalies can be deciphered as follows:

- Locating manual segmentation points of the path of the cursive area;
- Uncertainty of the hand of the calligrapher;
- Pretreatment of the path cursive area;
- Quality of materials.

*3) Reduction of the Cursive Area:*

Two points of cutting are to be determined:

- The output of the first letter;
- The input of the second letter.

For this, we decompose the trace of cursive area in sub-segments. We choose the points of curvature as indicative limits of these sub-segments.

The output of the first letter has to be after the input of the second. The area bounded by both points is the reduced and the common area by the two neighboring letters.

*4) Duplicate then Merge:*

- Duplicate the common area;
- Merge each area to two-letter graphemes.

For example, (see Fig. 18).

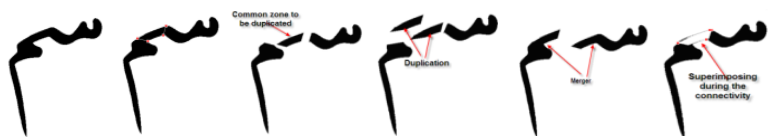

Fig. 18　　Stages of cutting

VIII. APPLICATIONS

*A. Digital Fonts of Calligraphic Glyphs*

The superimposing of the graphs obtained after the cutting, for the same letter and in the same context, allows identifying the uncertainties of the hand of a calligrapher during the graphic production. Looked for by an average shape is an obligation (bond).

*1) Variability of the Shape:*

By stacking the graphs obtained from the same letter and in the same context, it is possible to study the variability of the forms. This variability can be caused by an uncertainty of the hand, the calligrapher's behavior and equipment used.

*2) Average Shape:*

To look for a shape average cut graphs, the forms must be stacked and normalized by way of the Procrustes method. This method allows obtaining information on the distance between two forms. The first stage in this method is to look for, in the shape to be studied, certain number of points considered as reference points or points of interests which can summarize the studied shape. The aim is to obtain a similar placement and size, by minimizing a measure of difference in shape called the Procrustes distance between the objects [20].

*3) Rendering:*

In a font built from any calligrapher work, we obtain a record of quality in the reproduction of the same words cut.

When the context is varied, two problems may arise when composing new words:

- A variety of thickness (see Fig. 19);
- A non-regularity of the handwriting area.

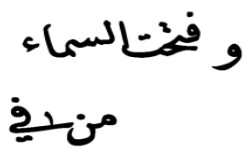

Fig. 19　　Variety of thickness

*4) Proposals:*

Reducing the number of alternatives is a requirement for technical font formats with limited capacity. It is also a need for a finite and manageable number of allographs to present in a font. In order to do so, we should:

- Normalize the inputs and outputs.
- Use the items as separate characters and Arabic letters on issues such as composite characters.
- Make the positioning of points by the composition system.
- Minimize the number of diacritics, consider diacritical Shadda as "diacritical composite".
- Have alternate cursive areas, and merge with parts of letters containing the radicals must be a task of the system.
- Consider thick endings unified glyphs.

*B. Classification of Allographs*

Allographs classification is important for classifying contexts and extracting calligraphic rules from handwritten manuscripts. It can be used also in a field of pattern recognition. In another work, we try to use artificial neural network for resolving the problem.

*C. Location of the Position of a Diacritic*

*1) Reference:*

For each letter, we define a benchmark consisting of:

- The baseline as x-axis and another that is perpendicular, tangent to the input of the base letter.
- The two-point unit consists of the dimensions of width and height of the base letter.

The projection of diacritic on the baseline is a segment. The middle of this segment is the orthogonal projection of a point of diacritic, which qualifies as extremities to measure different distances from diacritic to the axes (coordinates) (see Fig. 20).

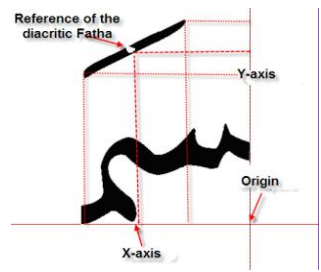

Fig. 20    Landmark for Fatha

*2) Interests:*

The positioning of diacritics with sizes suitable to the context is another issue against a digital reproduction authentic enough of Arabic calligraphy. The localization of diacritics by coordinates expressed in terms of the dimensions of length and width of the segmented basic letter, retrieves information on the positioning of diacritics and facilitates us develop a model of the diacritised Arabic text. This will be the subject of another study [21].

IX. CONCLUSION

The proposed segmentation procedure allows obtaining segments respecting the design of the Arabic letter, such as it was drawn by the calligrapher. It minimizes the number of information lost during the process of word cutting. However, the implementation for a digital production of font with calligraphic glyphs collides with the limits of the possibilities offered by font format or by rendering engines which correspond to it. The graphic segmentation of the Arabic word is a basic stage to revise the contextual analysis of the Arabic text and the strategies for positioning and sizing Arabic diacritical marks.

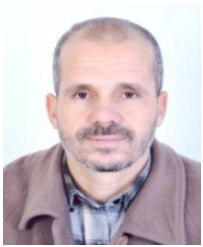

**Mohamed Hssini** is a Ph.D. student in Department of Computer Science at Cadi Ayyad University. He is a member of *multilingual scientific e-document processing* team. His current main research interest is multilingual typography, especially the publishing of Arabic e-documents while observing the strict rules of Arabic calligraphy.

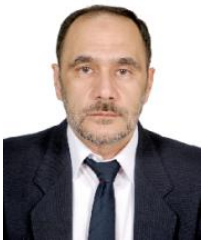

**Azzeddine Lazrek** is full Professor in *Computer Science* at Cadi Ayyad University in Marrakesh. He holds a *PhD* degree in *Computer Science* from Lorraine Polytechnic National Institute in France since 1988, in addition to a *State Doctorate* Morocco since 2002. Prof. Lazrek works on *Communication multilingual multimedia e-documents in the digital* area. His areas of interest include: multimedia information processing and its applications, particularly, electronic publishing, digital typography, Arabic processing and the history of science. He is in charge of the research team *Information Systems and Communication Networks* and the research group *Multilingual scientific e-document processing*. He is an *Invited Expert* at W3C. He leads a *multilingual e-document composition* project with other international organizations.